\RequirePackage{fix-cm}
\documentclass{article}      
\usepackage[dvipdfmx]{graphicx}
\usepackage{amsmath,amssymb}
\begin{document}

\title{Model of continuous random cascade processes in financial markets
}



\date{}
\maketitle

\centerline{ {\bf Jun-ichi Maskawa}$^{\rm a}$\footnote{\label{q}Jun-ichi Maskawa, 
Department of Economics, Seijo University
Address: 6-1-20 Seijo, Setagaya-ku, Tokyo 157-8511, Japan
Tel.: +81-3-3482-5938, Fax: +81-3-3482-3660
E-mail: maskawa@seijo.ac.jp }
 ,  {\bf Koji Kuroda}$^{\rm b}$}

\vspace{10pt}

\centerline{$^{\rm a}$Department of Economics, Seijo University} \vspace{10pt}

\centerline{$^{\rm b}$Graduate School of Integrated Basic Sciences, Nihon University}

\vspace{10pt}

\begin{abstract}
This article present a continuous cascade model of volatility formulated as a stochastic differential equation. Two independent Brownian motions are introduced as random sources triggering the volatility cascade. One multiplicatively combines with volatility; the other does so additively. Assuming that the latter acts perturbatively on the system, then the model parameters are estimated by application to an actual stock price time series. Numerical calculation of the Fokker--Planck equation derived from the stochastic differential equation is conducted using the estimated values of parameters. The results reproduce the pdf of the empirical volatility, the multifractality of the time series, and other empirical facts.
\end{abstract}

\section{Introduction}
In financial time series, past coarse-grained measures of volatility correlate better to future fine-scale volatility than the reverse process does. Such a causal structure of financial time series was first reported by M\"uller et al. \cite{muller1997}. Since then, the causal structure between time scales, the flow of information from a long-term to a short-term scale, was investigated empirically in financial markets; it has been supported by multiple studies \cite{arneodo1998a, lynch2003} as a stylized fact of financial time series \cite{cont2001}. The asymmetric flow of information resembles an energy cascade found in conditions of turbulence. In a developed turbulent flow, the energy cascades from the macroscopic spatial scale, where energy is injected from the outside, to the microscopic spatial scale, where energy is dissipated as heat \cite{richardson1922, kolmogorov1941, kolmogorov1962, mandelbrot1974, frisch1997}. Gashghaie et al. investigated details of the self-similar transformation rule of the probability density function of price fluctuations and the nonlinear scaling law of the structure function (n-th moment of fluctuations), signifying the multifractality of the time series, in their study of the time series of foreign exchange. They pointed out the similarity of price changes in the financial time series to the velocity difference between two spatial points in turbulence \cite{ghash1996,schmitt1999}. The intermittency in turbulence is a phenomenon by which a laminar flow is interrupted by irregular bursts that occur suddenly. Such intermittency, which is frequently encountered in heterogeneous complex systems, is well known in financial markets as volatility clustering \cite{cont2001, bouchaud_potter}. Intermittency at each time scale produces a characteristic hierarchical structure designated as multifractality \cite{mandelbrot1974, frisch1997}.

In the developed turbulence, the process by which mechanically generated vortices on a macro scale deform and destabilize according to the Navier--Stokes equation and then split into smaller vortices is regarded as an energy cascade. A similar idea of modeling multifractal time series by a recursive random multiplication process from a coarse-grained scale to a microscopic scale has offered an attractive means of describing financial time series \cite{abm1998,breymann2000}. Chen et al. verified the statistics of multiplier factors in the random multiplication process of turbulent flow by empirical studies using measured data and numerical experiments of Navier--Stokes equations \cite{chen2003}. Results show that the multiplier factors connecting two adjacent layers follow a Cauchy distribution in which all moments diverge, and show that they are not independent. They show strongly negative correlation between the multiplier factors of adjacent layers. The authors verified the statistics of multipliers calculated backward from actual stock price fluctuations, finding a Cauchy distribution of multiplier factors and also the strongly negative correlation between the multiplier factors in financial markets. Results show that the discrete cascade model using the random multiplication process did not reproduce the statistical property of the multiplier factors.  Therefore, as an alternative model, a discrete random multiplicative cascade process with additional additive stochastic process \cite{jim2000,jim2007,maskawa2018}, or a model formulated as the Fokker--Planck equation considering the cascade process as a continuous Markov process \cite{friedrich1997,renner2001,renner2001b,siefert2007,reinke2018} was proposed. Those models have been applied to stock market or foreign exchange market data, yielding empirical results including the statistics of multipliers.

This study examines a continuous cascade model of volatility formulated as a stochastic differential equation including two independent modes of Brownian motion: one has multiplicative coupling with volatility; the other has additive coupling as in the discrete random multiplicative cascade process with additional additive stochastic processes described above. The model parameters are estimated by its application to the stock price time series. Numerical calculation of the Fokker--Planck equation derived from the stochastic differential equation is conducted using the estimated values of parameters resulting in successful reproduction of the pdf of the empirical volatility and the multifractality of the time series.
\vspace{1cm}
\section{Materials and Methods}

\subsection{Continuous random cascade model}
\vspace{1cm}
\subsubsection{Stochastic differential equation}
These analyses examine the following wavelet transform of the variation of the logarithmic stock price denoted by $Z(t)=\log S(t)/S(0),~t\in[0,L]$:
\begin{equation}
	W_\psi Z[u,s]=\int^{+\infty}_{-\infty}Z(t)\frac{1}{s}\psi^*(\frac{t-u}{s})dt,~u\in[0,L], 
\label{swt}
\end{equation}
where function $\psi$ is designated as the analyzing wavelet. When using the delta function $\psi(t)=\delta(t-1)-\delta(t)$ as the analyzing wavelet, the wavelet transform $W_\psi Z[u,s]=Z(u+s)-Z(u)$ is exactly the logarithmic return of the period $s$. Here we use the second derivative of the Gaussian functions as
\begin{equation}
	\psi(t)=\frac{d^2}{dt^2}(e^{-\frac{t^2}{2}})=(t^2-1)e^{-\frac{t^2}{2}}.
\label{mexican}
\end{equation}
In general, by using the $n$-th derivative of the function having asymptotic fast decay as the analyzing wavelet, one can remove the local trend of $m$-th order ($m \le n-1$) because the function is orthogonal to $m$-th order polynomials. For the second derivative of the Gaussian functions, the linear trends of $Z(t)$ with scale $s$ have been eliminated in the wavelet transform $W_\psi Z[u,s]$.

The quantity used herein is the absolute value of the wavelet transform $x(\lambda)=| W_\psi Z[t_0, s(\lambda)]|$ for arbitrary $t_0$, where we use the variable $\lambda=\log L/s$. Quantity $x(\lambda)$ is thought to be a generalization of empirical volatility, whereas wavelet transform $W_\psi Z[u,s]$ is exactly the absolute value of logarithmic return when we use $\psi(t)=\delta(t-1)-\delta(t)$.

The following stochastic equation is used to start.
\begin{equation}
x(\lambda+d \lambda)=x(\lambda) \cdot e^{\sigma dB(\lambda)+
\mu d \lambda}~~( d\lambda>0) 
\label{cascadeeq}
\end{equation}
In that equation, $B(\lambda)$ represents the Brownian motion. Equation (\ref{cascadeeq}) expresses that the value of the quantity $x(\lambda+d \lambda)$ at scale $\lambda+d \lambda$ is obtained stochastically from $x(\lambda)$ at just a slightly larger scale $\lambda$ by multiplying the stochastic variable $W(\lambda, \lambda+ d\lambda)= e^{ \sigma dB(\lambda) + \mu  d \lambda } $ .  The stochastic multiplier $W(\lambda, \lambda+ d\lambda)$ follows a logarithmic normal distribution $LN(\mu d \lambda, \sigma^2 d \lambda)$ because $dB(\lambda ) \sim N(0, d \lambda)$ .
One can derive the following stochastic differential equation, using $dB(\lambda)^2=d\lambda$ as
\begin{align}
dx(\lambda)&~=x(\lambda+d\lambda)-x(\lambda)
\nonumber\\
&~=x(\lambda) \cdot \Bigl( e^{\sigma dB(\lambda)+\mu d\lambda}-1\bigr)\nonumber\\
&~=x(\lambda) \cdot  \Bigl( \sigma dB(\lambda)+ (\mu+\frac12 \sigma^2)d\lambda \Bigr).
\label{langevin1}
\end{align}

The solution is obtained easily using Ito's formula as \cite{gardiner}
\begin{equation}
x(\lambda)=x(0) \cdot  e^{\sigma B(\lambda)+\mu \lambda}.
\label{solution1}
\end{equation}

The power law behavior of the $q$-th moment $E[x(\lambda)^q]$ ($q$-th structure function) as a function of scale $s$ is proved by the solution (\ref{solution1}) as the following.
\begin{align}
E[x(\lambda)^q]&~=E[x(0)^q] \exp \bigl\{\mu \lambda q + \frac12 \sigma^2 \lambda q^2 \Bigr\}
\nonumber\\
&~=E[x(0)^q] \bigl( \frac{s}{L} \bigr)^{-\mu q -\frac12 \sigma^2 q^2}
\label{xq}
\end{align}

The multifractality of signal $Z(t)$ for which the wavelet transform follows the stochastic equation (\ref{langevin1}) is verified because the scaling exponent $\tau(q)=-\mu  q - \frac12 \sigma^2 q^2-1$ is a convex upward nonlinear function. However, in this model, the stochastic multiplier $W(\lambda_2, \lambda_1)=x(\lambda_2)/x(\lambda_1)~(\lambda_2 \le \lambda_1)$ linking two scales follows the logarithmic normal distribution $LN(\mu (\lambda_1-\lambda_2), \sigma^2 (\lambda_1-\lambda_2))$. It is independent of the multiplier $W(\lambda_3, \lambda_2)~(\lambda_3 \le \lambda_2)$ linking two adjacent scales. That result is contrary to the empirical results described in the Introduction.

We introduce an additional additive stochastic process as we have done into the discrete cascade model.
We first consider the following stochastic differential equation.
\begin{equation}
dx(\lambda)= x(\lambda)\cdot(-\gamma_Md\lambda +\sigma_MdB_M(\lambda))+a_A(\lambda)d\lambda+b_A(\lambda)dB_A(\lambda)
\label{langevin2}
\end{equation}
The equation is produced on the assumption that Brownian motions $dB_M(\lambda)$ and $dB_A(\lambda)$ are mutually independent. The first two terms correspond to equation (\ref{langevin1}). The origin of those random sources triggering volatility cascade in financial markets remains unclear.

To solve the stochastic differential equation (\ref{langevin2}), we consider the following stochastic differential equation of
\begin{equation}
dw(\lambda)=w(\lambda) \cdot (-\gamma_Md\lambda +\sigma_MdB_M(\lambda))
\label{langevin3}
\end{equation}
which is the same as (\ref{langevin1}). Using the solution of (\ref{langevin3})
\begin{equation}
w(\lambda)= w(0)\cdot \exp \bigl\{-\bigl( \gamma_M+\frac12 \sigma_M^2 \bigr) \lambda+\sigma_MB_M (\lambda) \Bigr\},
\label{solution3}
\end{equation}
the solution of (\ref{langevin2}) is expressed as shown below:
 \begin{equation}
x(\lambda)=w(\lambda)\cdot
\Bigl(
\int_0^\lambda \frac{a_A(u)}{w(u)} du
+ \int_0^\lambda
\frac{b_A(u)}{w(u)} dB_A(u)+ \frac{x(0)}{w(0)} \Bigr).
\label{solution2}
\end{equation}

\vspace{1cm}
\subsubsection{Statistics of multipliers}
We have mentioned the statistics of multipliers in the Introduction:
\begin{enumerate}
\item[(1)] The stochastic multiplier $W(\lambda_2, \lambda_1)=x(\lambda_2)/x(\lambda_1)~(\lambda_2 \le \lambda_1)$ linking two different scales follows a Cauchy distribution.
\item[(2)] When considering the three scales $\lambda_1 < \lambda_2 <\lambda_3~(s_1 > s_2 >s_3)$, the adjacent multipliers $W(\lambda_2, \lambda_1)=x(\lambda_2)/x(\lambda_1)$ and $W(\lambda_3, \lambda_2)=x(\lambda_3)/x(\lambda_2)$ show strongly negative correlation.
\end{enumerate}

Here we show property (1) and infer the existence of correlation between adjacent multipliers under some reasonable approximations.
Parameter $\sigma_M$ is an important model parameter for the signal to have multifractality. 
As presented in a later section, in spite of the importance, the value of the parameter $\sigma_M^2$ is tiny, about $0.02--0.03$ in stock markets, irrespective of the stock issue.
To specifically examined the role of additional stochastic processes, we investigate the 0-th order approximation of small  $\sigma_M$.
When setting $\sigma_M=0$, the solution (\ref{solution2}) becomes
\begin{equation}
x(\lambda)=
\Bigl(
\int_{\lambda_0}^\lambda a_A(u)e^{\gamma_M(u-\lambda_0)} du
+\int_{\lambda_0}^\lambda
b_A(u)e^{\gamma_M(u-\lambda_0)}dB_A(u)+ x(\lambda_0) \Bigr)
\label{solution3}
\end{equation}

Therefore, the difference $\Delta x(\lambda,\lambda_0)=x(\lambda)-x(\lambda_0)$ follows a normal distribution \\
$N(\int_{\lambda}^{\lambda_0 }a_A(u)e^{\gamma_M(u-\lambda_0)} du, \int_{\lambda_0}^\lambda (b_A(u))^2e^{2\gamma_M(u-\lambda_0)}du)$. If one simply assumes that $x(\lambda_0)$ follows a normal distribution, then the ratio $\Delta x(\lambda_0,\lambda_1)/x(\lambda_0)$ of two independent stochastic variables following normal distributions follows a Cauchy distribution. So $x(\lambda_1)/x(\lambda_0)$ is the same.  

By defining the differences $\Delta x(\lambda_2,\lambda_1)=x(\lambda_2)-x(\lambda_1)$  and $\Delta x(\lambda_3,\lambda_2)=x(\lambda_3)-x(\lambda_2)$ for the three scales $\lambda_1 < \lambda_2 <\lambda_3$, it is readily apparent that $W_1=x(\lambda_2)/x(\lambda_1)=1+\Delta x(\lambda_2,\lambda_1)/x(\lambda_1)$ and $W_2=x(\lambda_3)/x(\lambda_2)=1+\Delta x(\lambda_3,\lambda_2)/x(\lambda_2)$ show correlation. In this framework, it was difficult to show that they have strongly negative correlation. Those statistics of multipliers have also been considered in earlier works by Siefert and Peinke \cite{siefert2007}. The same result can be shown using a Fokker--Planck equation under some approximations. In a later section, we show a similar Fokker--Planck equation derived from the stochastic differential equation (\ref{langevin2}).

\vspace{1cm}
\subsubsection{Relation to discrete random cascade model}
Assuming that $\Delta \lambda$ is sufficiently small, then when we use the following approximation of Ito's stochastic integration\cite{gardiner} as
\begin{equation}
\int_{\lambda}^{\lambda+\Delta\lambda} \frac{b_A(u)}{w(u)}dB_A(u)\sim \frac{b_A(\lambda)}{w(t)}(B_A(\lambda+\Delta\lambda)-B_A(\lambda)),
\label{}
\end{equation}
we obtain the discrete random cascade equation as
\begin{align}
&x(\lambda+\Delta\lambda)\nonumber\\
&=W_M(\lambda,\lambda+\Delta\lambda)\cdot(x(\lambda)+a_A(\lambda)\Delta\lambda +b_A(\lambda)(B_A(\lambda+\Delta\lambda)-B_A(\lambda)),
\label{model2}
\end{align}
where $W_M(\lambda,\lambda+\Delta\lambda)=e^{ - (\gamma_M+\frac 12\sigma_M^2) \Delta\lambda  +\sigma_M ( B_M(\lambda+ \Delta\lambda) -B_M(\lambda))} $. The conditional expectation value of the square of $x(\lambda+\Delta\lambda)$, as the function of $x^2(\lambda)$,
\begin{equation}
E(x^2(\lambda+\Delta\lambda)|x(\lambda))=e^{(2\mu_M+2 \sigma_M^2)\Delta\lambda}(x^2(\lambda)+(2a_A(\lambda)x(\lambda)+b_A^2(\lambda)),\Delta\lambda))
\label{e2}
\end{equation}
shows that deviation of the quadratic curve from the origin results from the parameter $b_A(\lambda)$, as demonstrated from an empirical study in \cite{maskawa2018}.

\vspace{1cm}
\subsubsection{Constraint condition from the pdf of $x(\lambda)$}
\begin{figure}[h!]
\begin{center}
\includegraphics[width=10cm]{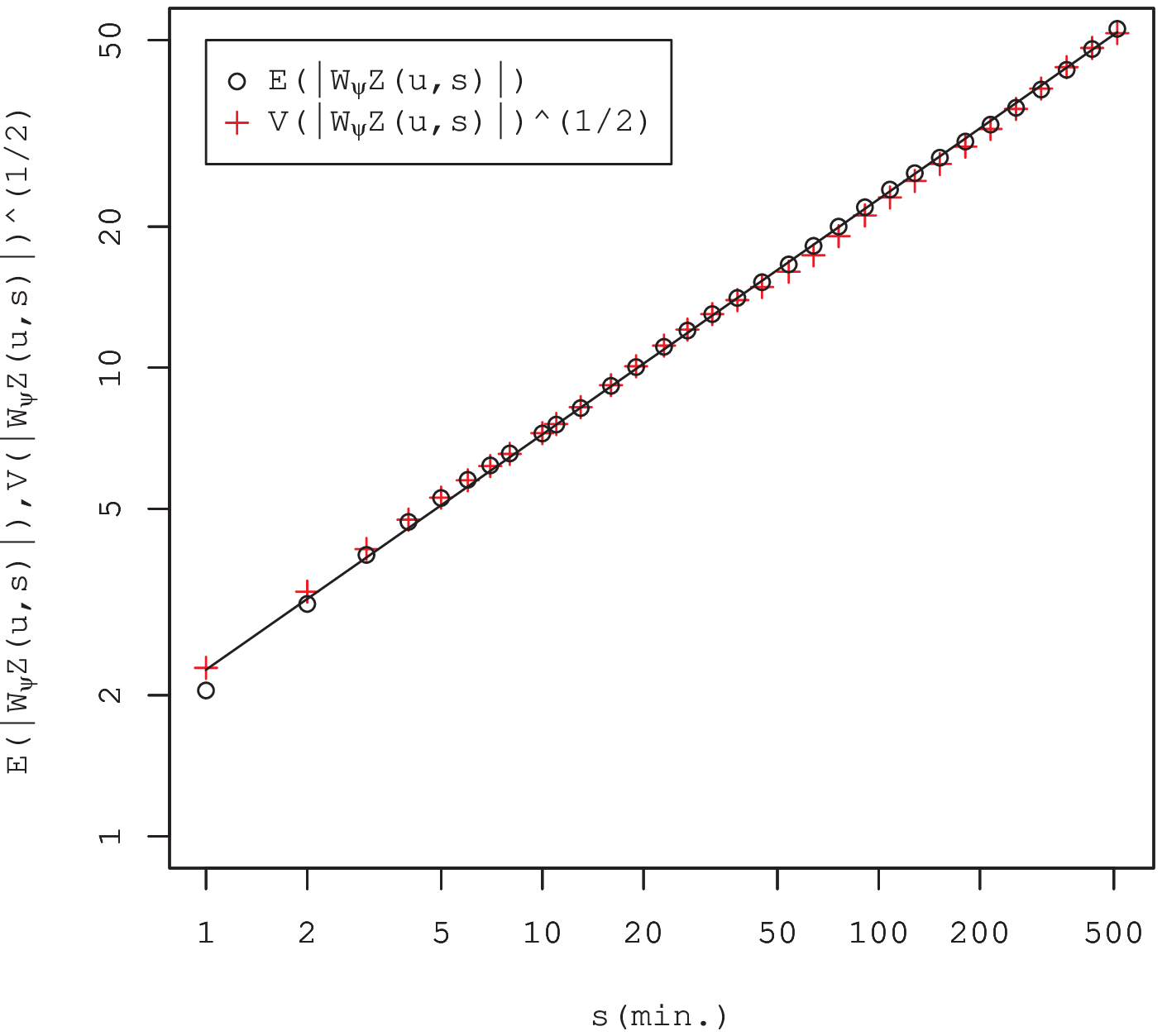}
\end{center}
\caption{Scaling properties of $E(|W_\psi Z[., s(\lambda)]|)$ and $V(|W_\psi Z[., s(\lambda)]|)^{1/2}$. The expected value almost perfectly coincides with the standard deviation at all scales. The solid line represents the least-squares fit to the power law function, $2.27s^{0.5}$.}
\label{fig1}
\end{figure}

A remarkable feature of the probability density function (pdf) of the quantity $x(\lambda)=|W_\psi Z[., s(\lambda)]|$ is the coincidence of the expected value $E(|W_\psi Z[., s(\lambda)]|)$ with standard deviation $V(|W_\psi Z[., s(\lambda)]|)^{1/2}$, as shown in Fig. \ref{fig1} for the data examined in this study (see also Fig. \ref{fig10} for the pdf of $x(\lambda)$). It indicates the constraint condition as
\begin{equation}
a_A(\lambda)E(x(\lambda))=b_A^2(\lambda).
\label{constraint}
\end{equation}
Derivation of the constraint condition (\ref{constraint}) is given in the Appendix below.

The additional additive stochastic process in the model (\ref{langevin2}) is expected to be a small perturbation to the basic model (\ref{langevin1}) to avoid violating multifractality. We also impose the following condition for all scales s:
\begin{equation}
\frac{a_A(s)}{E(|W_\psi Z[.,s]|)}<<1,~~\frac{b_A(s)}{E(|W_\psi Z[.,s]|)}<<1.
\label{assumption1}
\end{equation}
The power law scaling shown in Fig. \ref{fig1}, 
\begin{equation}
E(|W_\psi Z[.,s]|)\sim s^{0.5},
\label{EW}
\end{equation}
and condition (\ref{assumption1}) show the following constraint condition:
\begin{equation}
a_A(s)\sim s^{0.5},~~b_A(s)\sim s^{0.5}.
\label{assumption2}
\end{equation}
Inserting (\ref{assumption2}) into (\ref{constraint}), we also have the equation
\begin{equation}
a_A(1)E(|W_\psi Z[.,1]|)=b_A^2(1).
\label{assumption3}
\end{equation}

\subsubsection{Appendix: Derivation of (\ref{constraint})}
We introduce some notation for simplification of the description:
\begin{equation}
E_1(\lambda)=E(x(\lambda)),~E_2(\lambda)=E(x^2(\lambda)),~\Delta B_\lambda=B_A(\lambda+d\lambda)-B_A(\lambda),~\mu_M=-(\gamma_M+\frac 12 \sigma_M^2).
\nonumber
\label{notation}
\end{equation}
From equation (\ref{model2}), we have
\begin{align}
E_1(\lambda+d\lambda)&=E(W_M(\lambda,\lambda+d\lambda)\cdot(x(\lambda)+a_A(\lambda)d\lambda+b_A(\lambda)\Delta B_\lambda))
\nonumber\\
&=E(W_M(\lambda,\lambda+d\lambda))E(x(\lambda)+a_A(\lambda)d\lambda+b_A(\lambda)\Delta B_\lambda))
\nonumber\\
&=E(W_M(\lambda,\lambda+d\lambda))(E_1(\lambda)+a_A(\lambda)d\lambda)
\nonumber\\
&=e^{(\mu_M+\frac 12 \sigma_M^2)d\lambda}(E_1(\lambda)+a_A(\lambda)d\lambda),
\nonumber
\end{align}
\begin{align}
E_2(\lambda+d\lambda)&=E(W_M^2(\lambda,\lambda+d\lambda)\cdot(x(\lambda)^2+(2a_A(\lambda)x(\lambda)+b_A^2(\lambda))d\lambda)
\nonumber\\
&=E(W_M^2(\lambda,\lambda+d\lambda))E(x(\lambda)^2+(2a_A(\lambda)x(\lambda)+b_A^2(\lambda))d\lambda)
\nonumber\\
&=E(W_M^2(\lambda,\lambda+d\lambda))(E_2(\lambda)+(2a_A(\lambda)E_1(\lambda)+b_A^2(\lambda))d\lambda))
\nonumber\\
&=e^{(2\mu_M+2 \sigma_M^2)d\lambda}(E_2(\lambda)+(2a_A(\lambda)E_1(\lambda)+b_A^2(\lambda))d\lambda)).
\nonumber
\end{align}
We also have
\begin{equation}
E_1^2(\lambda+d\lambda)=e^{(2\mu_M+\sigma_M^2)d\lambda}(E_1^2(\lambda)+2a_A(\lambda)E_1(\lambda)d\lambda).
\nonumber
\label{}
\end{equation}
Because of the coincidence of the expected value and the standard deviation, we have $E_2(\lambda)=2E_1^2(\lambda)$ and $E_2(\lambda+d\lambda)=2E_1^2(\lambda+d\lambda)$. Inserting those equalities, and using approximation $e^{\sigma_M^2d\lambda}=1$, we have the constraint condition (\ref{constraint}).

\vspace{1cm}
\subsection{Fokker--Planck equation}
We can derive the Fokker--Planck equation for the stochastic process $\{x(\lambda)\}$ expressed by the stochastic differential equation (\ref{langevin2}) as the following, which is the master equation that the density of the transition probability $p(x,\lambda |x_0,\lambda_0)$ follows\cite{gardiner}.
\begin{align}
&\frac{\partial}{\partial \lambda}p(x,\lambda |x_0,\lambda_0)
\nonumber\\
&\quad =  \Bigl[- \frac{\partial}{\partial x}D_1(\lambda,x)+\frac12 \frac{\partial^2}{\partial x^2}D_2(\lambda,x)\Bigr] p(x,\lambda |x_0,\lambda_0)
\label{FP}
\end{align}
Therein, the functions $D_1(\lambda,x)$ and $D_2(\lambda,x)$ are defined as 
\begin{align}
D_1(\lambda,x)&=a_A(\lambda)-\gamma_Mx,
\nonumber\\
D_2(\lambda,x)&=b_A(\lambda)^2+\sigma_M^2x^2.
\label{d1d2}
\end{align}
The k-th moment of the change $\delta x(\lambda)=x(\lambda+\delta \lambda)-x(\lambda)$ induced by the infinitesimal scale transformation $\delta \lambda$ is derived as shown below.
 \begin{align}
&E(\delta x^k|x(\lambda)=x)
\nonumber\\
&=\int_{-\infty}^{+\infty}(y-x)^kp(y,\lambda+\delta \lambda|x,\lambda)dy
\nonumber\\
&=\int_{-\infty}^{+\infty}(y-x)^k\bigl(p(y,\lambda|x,\lambda)+\delta \lambda \frac{\partial}{\partial \lambda '}p(y,\lambda '|x,\lambda)\bigr|_{\lambda '=\lambda}+O(\delta \lambda^2)\bigr)dy
\nonumber\\
&=\int_{-\infty}^{+\infty}(y-x)^k\bigl(\delta \lambda\bigr[ -\frac{\partial}{\partial y}D_1(\lambda ',y)+\frac 12 \frac{\partial^2}{\partial y^2}D_2(\lambda ',y)\bigr]p(y,\lambda '|x,\lambda)\bigr|_{\lambda '=\lambda}+O(\delta \lambda^2)\bigr)dy
\nonumber\\
&=\int_{-\infty}^{+\infty}\bigl(\delta \lambda \bigr[ k(y-x)^{k-1}D_1(\lambda,y)+\frac 12 k(\frac{\partial}{\partial y}(y-x)^{k-1})D_2(\lambda,y)\bigr]\delta(y-x)+O(\delta \lambda^2)\bigr)d
\nonumber\\
\label{K-M}
\end{align}
Therein, we used the identity $p(y, \lambda|x, \lambda)=\delta(y-x)$. Coefficients $D_1(\lambda,x)$ and $D_2(\lambda,x)$ show a relation to the first and second moments of $\delta x(\lambda)$ in the following way.
\begin{equation}
\lim_{\delta \lambda\rightarrow0}\frac{E(\delta x^k|x(\lambda)=x)}{\delta \lambda}=
\begin{cases}
D_k(\lambda,x)&k=1,2\\
0&others
\end{cases}
\label{KM}
\end{equation}
Coefficients $D_k$ are designated as Kramers--Moyal coefficients\cite{gardiner,risken}. We use equation (\ref{KM}) to estimate the function $a_A(\lambda)$ and $b_A(\lambda)$ and parameters $\gamma_M$ and $\sigma_M$.
To validate the model (\ref{langevin2}), it is necessary to confirm vanishing of the k-th moments for $3 \le k$ in the limit of $\delta \lambda\rightarrow0$. Renner et al. proposed almost identical equations with (\ref{FP}) in the literature \cite{renner2001,renner2001b}, in which they deal with the price change itself as an analogy of the velocity difference in turbulence \cite{friedrich1997}. They derived a Fokker--Planck equation as a result of their empirical studies using Kramers--Moyal expansion of the Chapman--Kolmogorov equation, regarding the process as a Markovian process.

\subsection{Empirical study}
\subsubsection{Data}
We analyze the normalized average of the logarithmic stock prices of the constituent issues of FTSE 100 index listed on the London Stock Exchange for November 2007 through January 2009, which includes the Lehman shock of 15 September 2008 and the market crash of 8 October 2008.

\subsubsection{Data processing}
First, we calculate the average deseasonalized return of each issue $\delta Z_i(t)=\log(S_i(t))-\log(S_i(t-\delta t))$, which describes the average change of the portfolio as
\begin{equation}
\delta Z(k\delta t)=\frac{1}{N_F}\sum_{i=1}^{N_F}\frac{\delta Z_i(k\delta t)-\mu_i}{\sigma_i},
\label{df}
\end{equation}
where $\mu_i$ and $\sigma_i$ respectively denote the average and the standard deviation of $\delta Z_i$ and where $N_F$ represents the number of constituent stock issues (stocks). The constituents of FTSE100 Index are updated frequently. We selected $N_F = 111$ stocks that remained listed on the London Stock Exchange throughout the period. Here, we set $\delta t=1$ (min) and examine the 1-min log-return.
We excluded the overnight price change and specifically examine the intraday evolutions of returns. To remove the effect of intraday U shape patterns of market activity from the time-series, the return was divided by the standard deviation of the corresponding time of the day for each issue $i$.
Then we cumulate $\delta Z(t)$ to obtain the path of process  $Z(k\delta t)~(k=1,\dots,L)$ (Fig. \ref{fig2}(A)) as
\begin{equation}
Z(k\delta t)=\sum_{k'=1}^k \delta Z(k'\delta t).
\label{f}
\end{equation}

\vspace{1cm}
\section{Results}
\vspace{1cm}
\subsubsection{Multifractal analysis}
First, we analyze the multifractal properties of the path $Z(t)$ using an approach with wavelet-based multifractal formalism proposed by Muzy, Bacry, and Arneodo \cite{muzy1993,bacry1993}. Initially, we define two mathematical terms. The H\"{o}lder exponent $\alpha(x_0)$ of a function $f(x)$ at $x_0$ is defined as the largest exponent such that there exists an $n$th-order polynomial $P_n(x)$ and constant $C$ that satisfy
\begin{equation}
|f(x)-P_n(x-x_0)| \le C|x-x_0|^{\alpha}
\label{holder},
\end{equation}
for $x$ in a neighborhood of $x_0$, characterizing the regularity of the function $f(x)$ at $x_0$. The singular spectrum $D(\alpha)$ is the Hausdorff dimension of the set where the H\"{o}lder exponent is equal to $\alpha$, as
\begin{equation}
D(\alpha)=dim_H\{x|\alpha(x)=\alpha\}.
\label{singular}
\end{equation}
For multifractal paths, the H\"{o}lder exponent $\alpha$ is distributed in a range; for paths of the Brownian motion, which are fractal, $D(0.5)=1$ and $D(\alpha)=0$ for $\alpha\ne0.5$.

\begin{figure}[h!]
\begin{center}
\includegraphics[width=10cm]{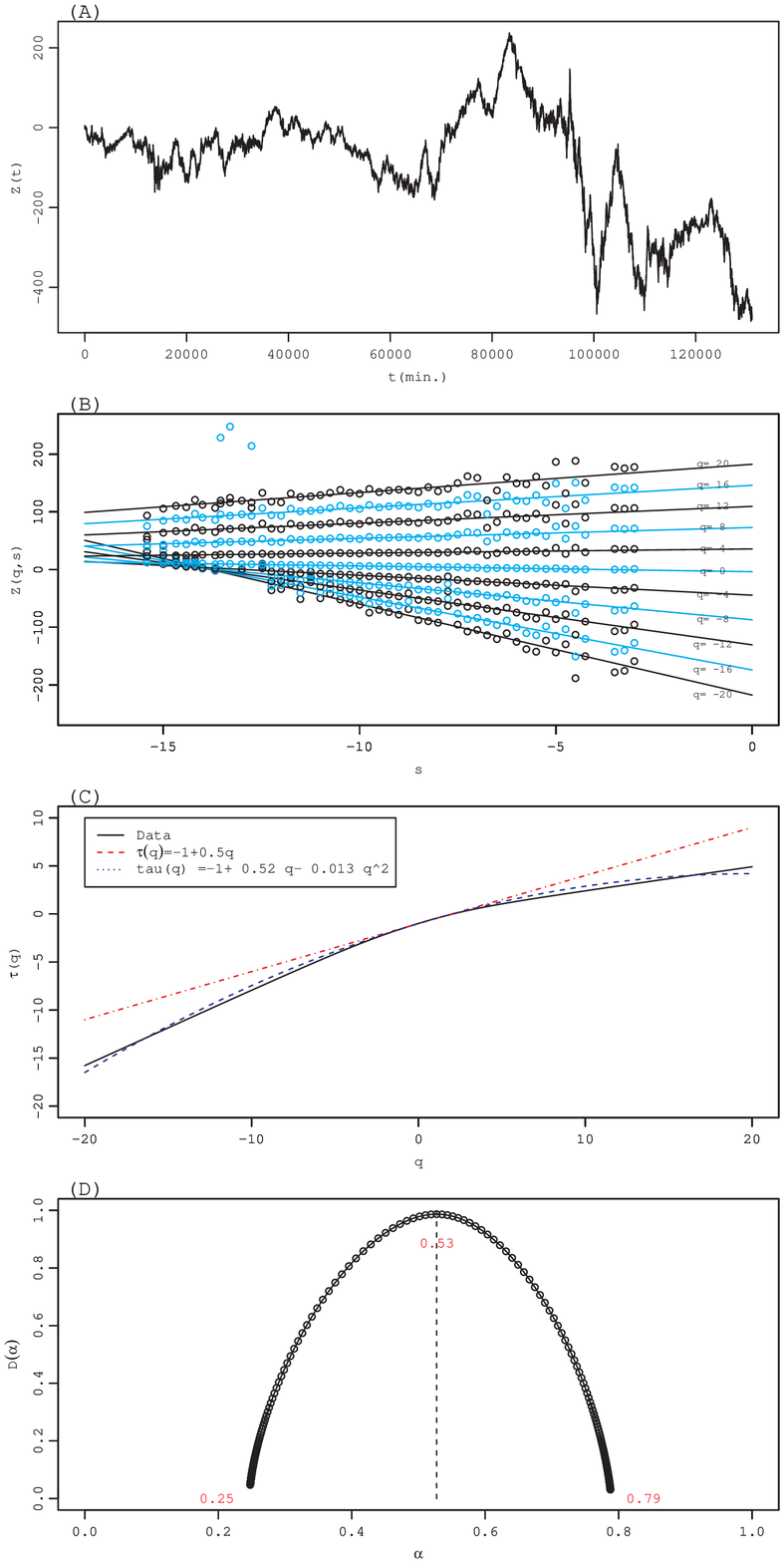}
\caption{Results of multifractal analysis (A)$Z(t)$ (B) $Z(q,s)$ for $q=-20,\dots,20$ ($\circ$) and regression lines (C) Scaling exponent $\tau(q)$ (solid line).  The dashed blue line is the least-squares fit to the quadratic function$\tau(q)=-1+0.52q-0.013q^2$. The dotted red line $\tau(q)=-1+0.5q$  corresponds to Brownian motion. (D) Singular spectrum $D(\alpha)$}
\label{fig2}
\end{center}
\end{figure}

Muzy, Bacry and Arneodo proposed the wavelet transform modulus maxima (WTMM) method based on continuous wavelet transform of function to calculate the singular spectrum $D(\alpha)$. We briefly sketch the WTMM method in the Appendix below. We calculate the partition function $Z(q,s)$ of the $q$-th moment of wavelet coefficients using equation (\ref{zqs}) for the path of our data. Results are presented in Fig. \ref{fig2}(B). The partition function $Z(q,s)$ for each order $q$ shows power law behavior in the range of scales $s/L<2^{-5}$. Exponents $\tau(q)$ are derived by the equation (\ref{tau}). Figure \ref{fig2}(C) shows that it is a convex function of $q$. Those results underscore the multifractality of the data path. The singular spectrum $D(\alpha)$ derived as the Legendre transformation of the function $\tau(q)$ by equation (\ref{D}) is a convex function that has compact support $[0.25,0.79]$ taking the peak at $\alpha=0.53$, as shown in Fig. \ref{fig2}(D).

\vspace{1cm}
\subsubsection{Appendix: WTMM method}
The WTMM method builds a partition function from the modulus maxima of the wavelet transform defined at each scale $s$ as the local maxima of $|W_{\psi}[f](x,s)|$ regarded as a function of x. Those maxima mutually connect across scales and form ridge lines designated as maxima lines.
The set $\mathcal{L}(s_0)$ is the set of all the maxima lines $l$ which satisfy
\begin{equation}
(x,s)\in l \Rightarrow s \le s_0,~\forall s \le s_0  \Rightarrow \exists(x,s)\in l.
\label{ml}
\end{equation}
The partition function is defined by the maxima lines as
\begin{equation}
Z(q,s)=\sum_{l\in \mathcal{L}(s)}(\sup_{(x,s')\in l}|W_{\psi}[x,s']|)^q.
\label{zqs}
\end{equation}
Assuming power-law behavior of the partition function
\begin{equation}
Z(q,s)\sim s^{\tau(q)},
\label{tau}
\end{equation}
one can define the exponents $\tau(q)$.
The singular spectrum $D(\alpha)$ can be computed using the Legendre transform of $\tau(q)$:
\begin{equation}
D(\alpha)=\min_q (q\alpha -\tau(q)).
\label{D}
\end{equation}

\vspace{1cm}
\subsubsection{Parameter estimations}
\begin{figure}[h!]
\begin{center}
\includegraphics[width=10cm]{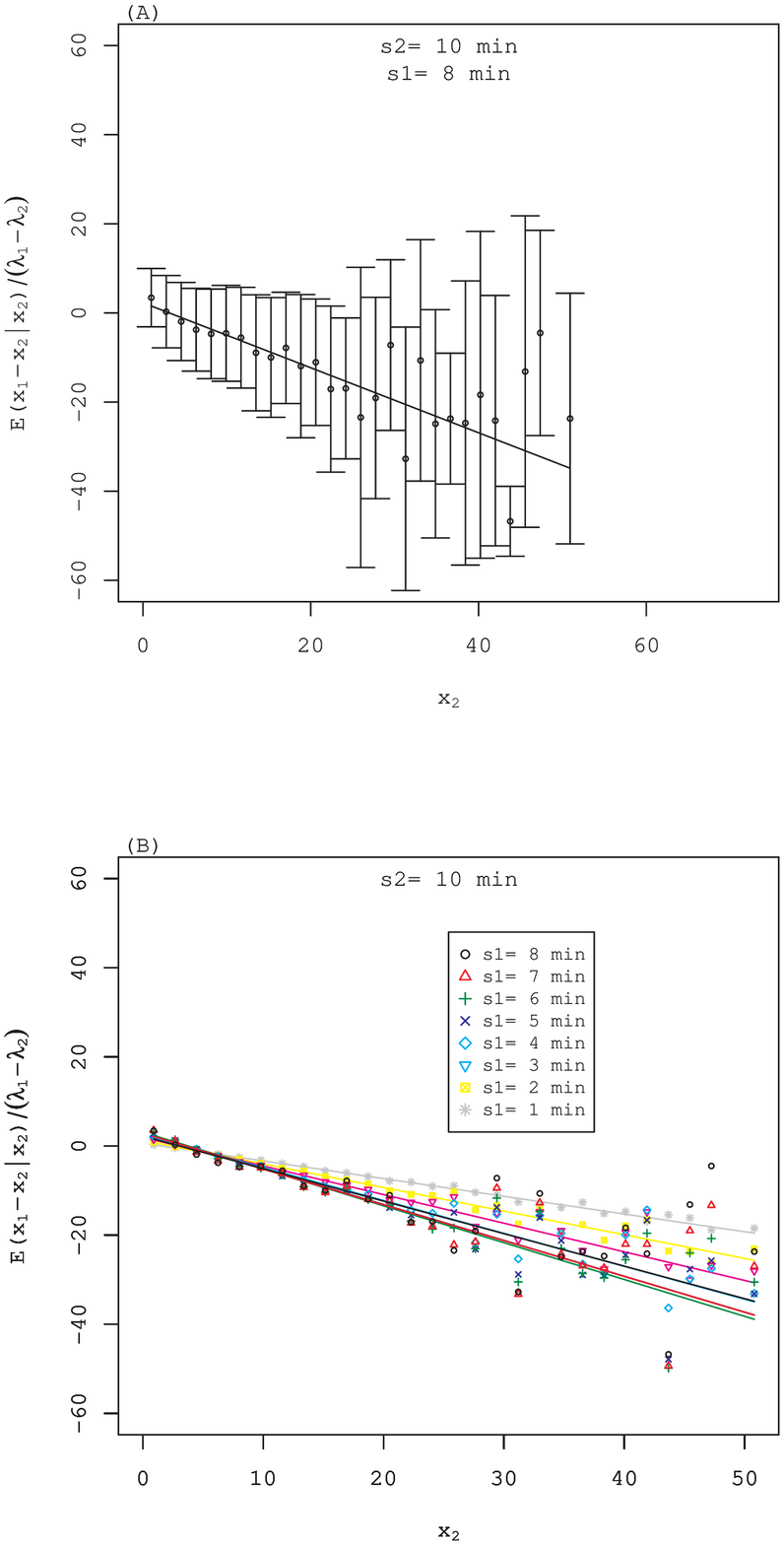}
\caption{Regression of $E(x_1-x_2|x_2=x)/(\lambda_1-\lambda_2)$. (A) The standard errors are denoted by an error bar. (B) Fitting is applied to various $s_1$ and $s_2$ combinations.}
\label{fig3}
\end{center}
\end{figure}

\begin{figure}[h!]
\begin{center}
\includegraphics[width=10cm]{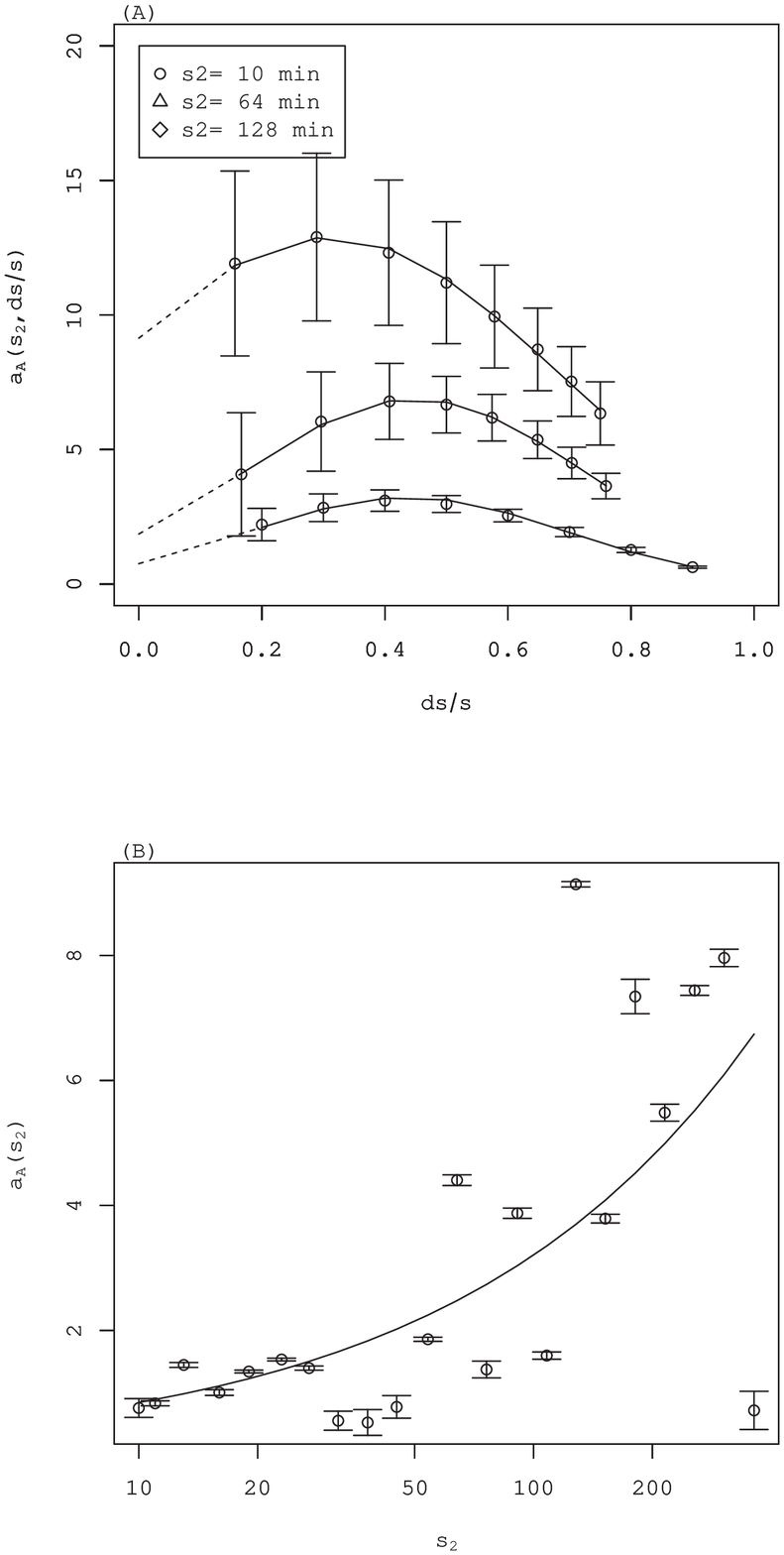}
\caption{Estimation of parameter $a_A(s)$. (A) Parameter $a_A(s_2,ds/s)$ obtained by the regressions shown in Fig. \ref{fig3} and nonlinear fitting $log(a_A(s,ds/s))=a+b(ds/s)+c(ds/s)^2$. The standard errors of regression (\ref{E1}) are denoted by an error bar. (B) $a_A(s)=exp(a)$ (see the text). The standard errors of nonlinear fittings are denoted by an error bar. The solid line shows the least-squares fit of $a_A(s)$ to the power law function.}\label{fig4}
\end{center}
\end{figure}

\begin{figure}[h!]
\begin{center}
\includegraphics[width=10cm]{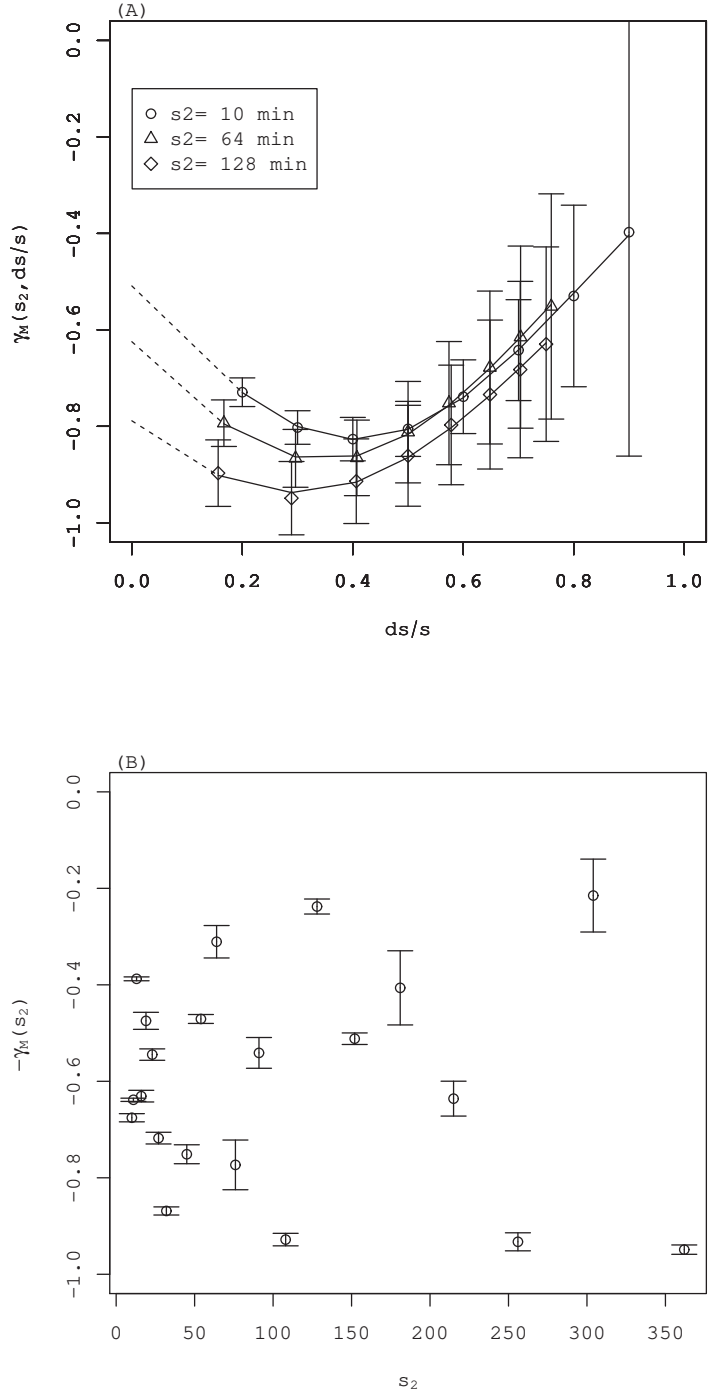}
\caption{Estimation of parameter $\gamma_M$. (A) Parameter $\gamma_M(s,ds/s)$ obtained by the regressions shown in Fig. \ref{fig3} and nonlinear fitting $log(\gamma_M(s))=a+b(ds/s)+c(ds/s)^2$. The standard errors of regression (\ref{E1}) are denoted by an error bar. (B) $\gamma_M(s)=exp(a)$ (see the text). Standard errors of nonlinear fittings are denoted by an error bar. }
\label{fig5}
\end{center}
\end{figure}

\vspace{1cm}
\paragraph{$a_A$ and $\gamma_M$}
Parameters $a_A(\lambda)$ and $\gamma_M$ are estimated by taking the limit $\lambda_1-\lambda_2\rightarrow 0$ of the first moment $E(x_1-x_2|x_2=x)/(\lambda_1-\lambda_2)$$(\ref{K-M})$.
The first moment $E(x_1-x_2|x_2=x)/(\lambda_1-\lambda_2)$ is fitted by a linear function as
\begin{equation}
\frac{E(x_1-x_2|x_2=x)}{d\lambda}=a_A(\lambda_2,d\lambda)-\gamma_M(\lambda_2,d\lambda)x,
\label{E1}
\end{equation}
where $d\lambda=\lambda_1-\lambda_2$. As shown in Fig. \ref{fig3}(a), the first moment is well fitted by a linear function. Fitting of this kind is applied to various $\lambda_1=\log(L/s1)$ and $\lambda_2=\log(L/s2)$ combinations (Fig. \ref{fig3}(B)). Taking the limit $d\lambda\rightarrow 0~(ds/s=(s_2-s_1)/s_2\rightarrow 0)$, one obtains $a_A(\lambda_2)=\lim_{d\lambda\rightarrow  0}a_A(\lambda_2,d\lambda)$($a_A(s_2)=\lim_{ds/s\rightarrow  0}a_A(s_2,ds/s)$) and $\gamma_M=\lim_{d\lambda\rightarrow  0}\gamma_M(\lambda_2,d\lambda)$($\gamma_M=\lim_{ds/s\rightarrow  0}\gamma_M(s_2,ds/s)$).
Fig. \ref{fig4}(A) presents examples of $a_A(s_2,ds/s)$ and nonlinear fittings by the function $log(a_A(s,ds/s))=a+b(ds/s)+c(ds/s)^2$. We estimate $a_A(s)$ by $a_A(s)=\exp(a)$ for each line. The result is presented in Fig. \ref{fig4}(B). The solid line is the least-squares fit $a_A(s)$ to a power law function as
\begin{equation}
\log(a_A(s))=-1.50(0.41)+0.58(0.11)\log s,
\label{a_A}
\end{equation}
where the standard errors are in parentheses. The estimated exponent $0.58(0.11)$ is consistent with the constraint condition (\ref{assumption2}) within the standard error. 
By a similar extrapolation $\log(\gamma_M(s,ds/s))=a+b(ds/s)+c(ds/s)^2$, we estimate $\gamma_M(s)=exp(a)$. Figure \ref{fig5}(A) presents examples of $\gamma_M(s_2,ds/s)$ and nonlinear fittings. We estimate $\gamma_M(s)$ by $\gamma_M(s)=\exp(a)$ for each line. The result is presented in Fig. \ref{fig5}(B). We estimate the parameter $\gamma_M$ by the average value weighted by the reciprocals of the standard errors as
\begin{equation}
\gamma_M=0.64(0.21),
\label{g_M}
\end{equation}
where the standard error is the value in the parenthesis.

\begin{figure}[h!]
\begin{center}
\includegraphics[width=10cm]{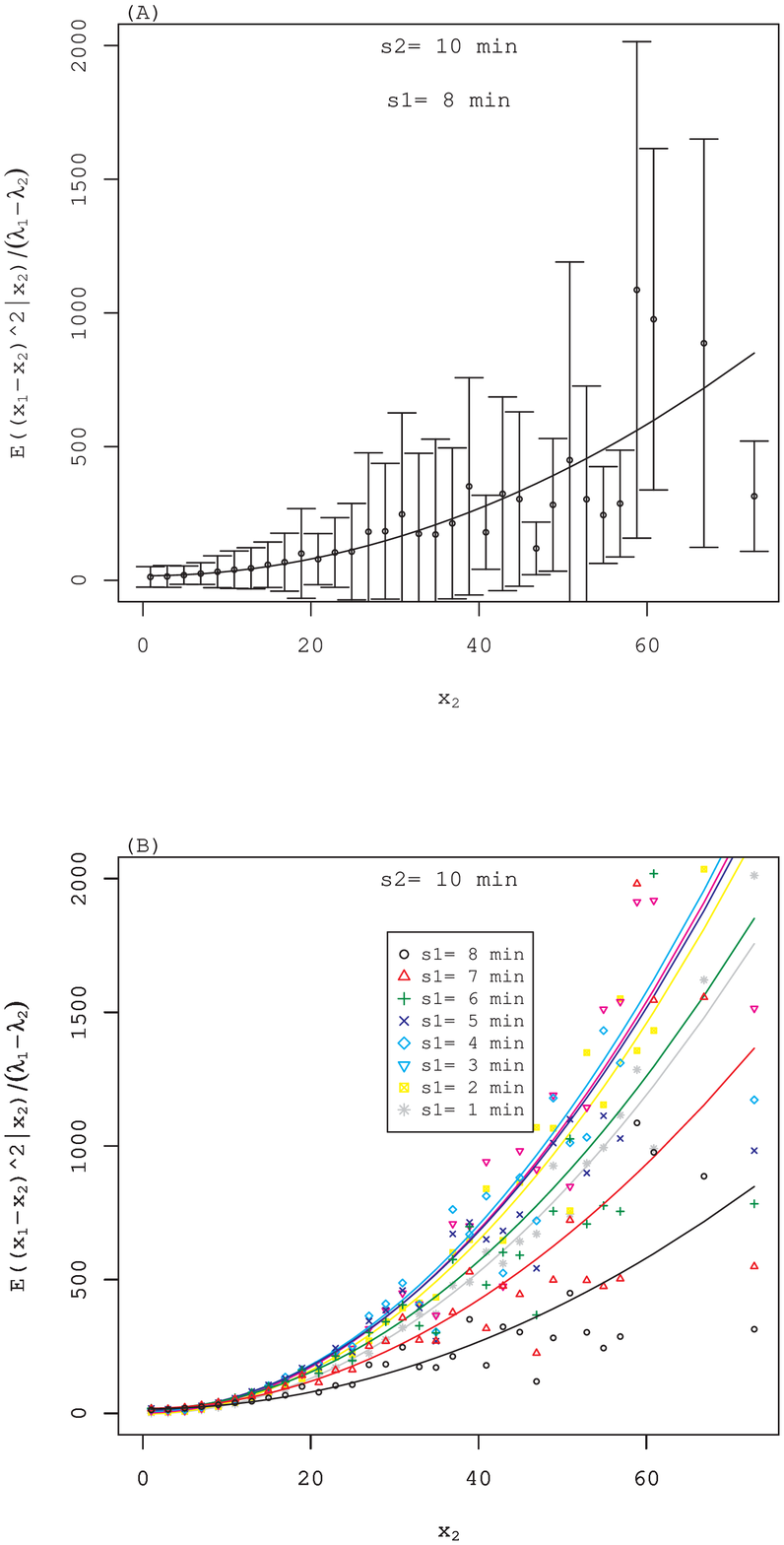}
\caption{Regression of $E((x_1-x_2)|x_2=x)/(\lambda_1-\lambda_2)$ against $x^2$. (A) Standard errors are denoted by an error bar. (B) Fitting is applied to various $s_1$ and $s_2$ combinations.}
\label{fig6}
\end{center}
\end{figure}

\begin{figure}[h!]
\begin{center}
\includegraphics[width=10cm]{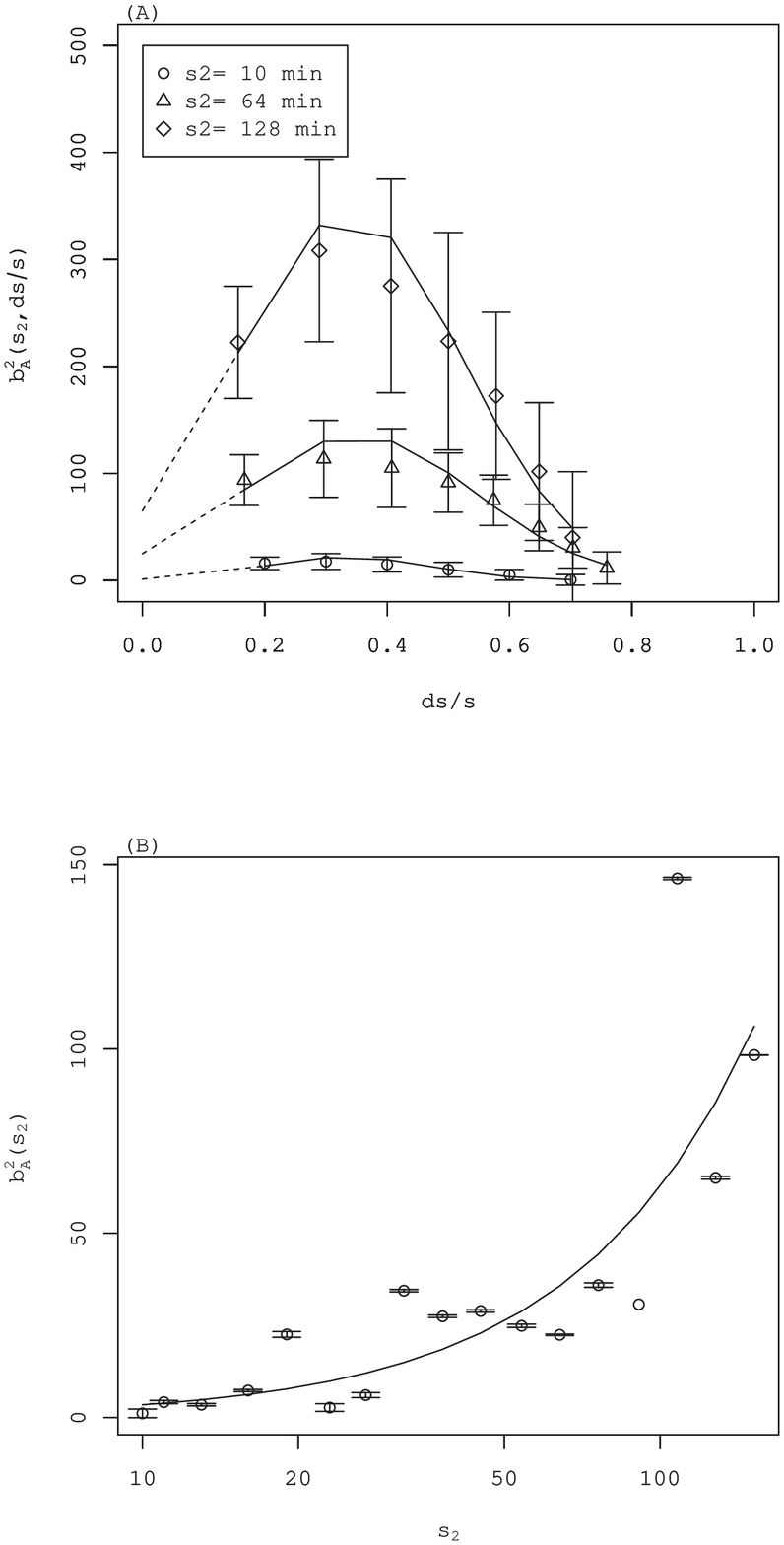}
\caption{Estimation of parameter $b_A(s)$. (A) Parameter $b_A(s_2,ds/s)$ obtained by the regressions shown in Fig. \ref{fig6} and nonlinear fitting $log(b_A^2(s,ds/s))=a+b(ds/s)+c(ds/s)^2$. The standard errors of the regression (\ref{E2}) are denoted by an error bar. (B) $b_A^2(s)=exp(a)$ (see the text). Standard errors of nonlinear fittings are denoted by an error bar. The solid line shows the least-squares fit of $b_A^2(s)$ to the power law function.}
\label{fig7}
\end{center}
\end{figure}

\begin{figure}[h!]
\begin{center}
\includegraphics[width=10cm]{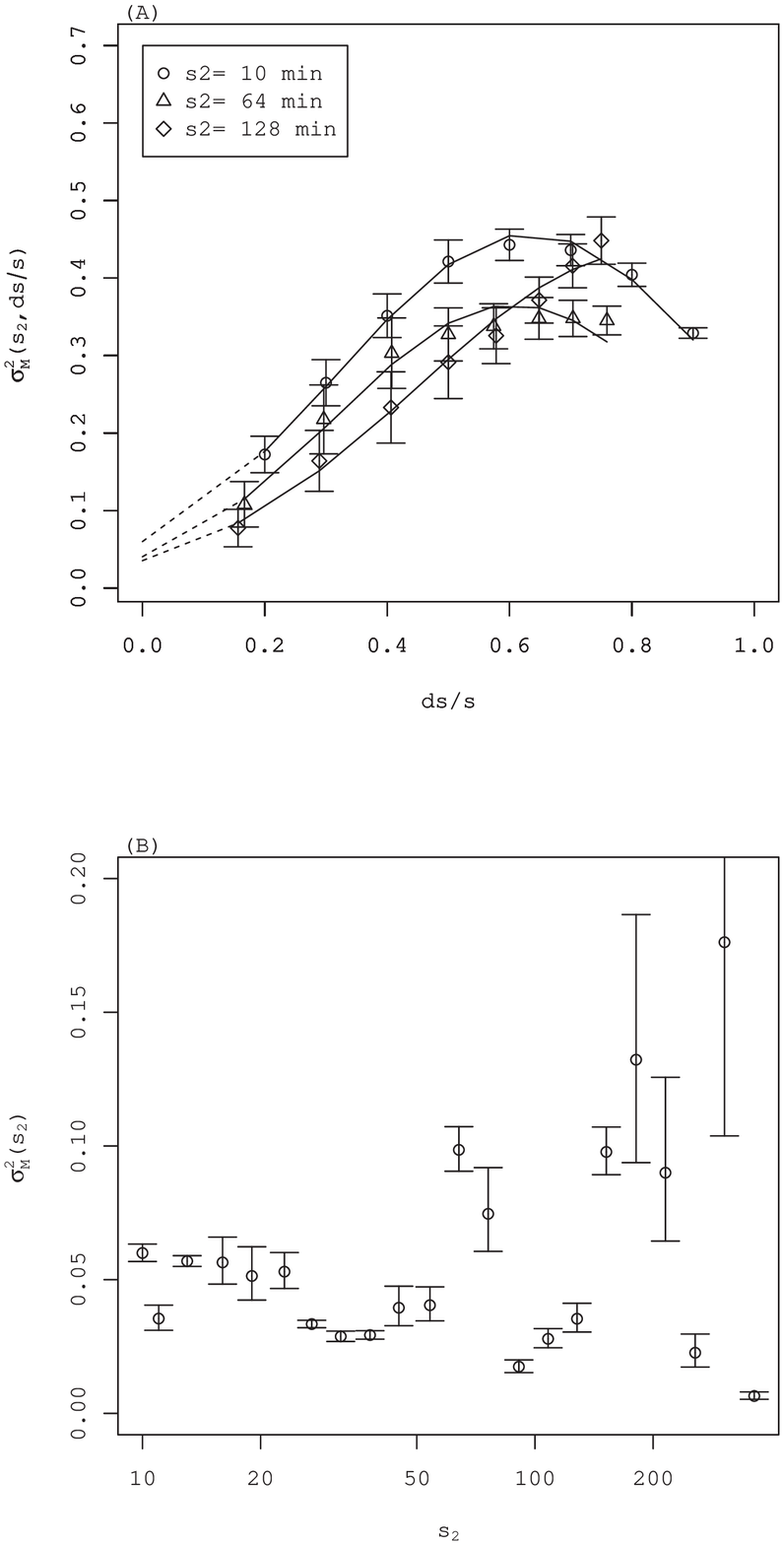}
\caption{Estimation of parameter $\sigma_M$. (A) Parameter $\sigma_M(s,ds/s)$ obtained by the regressions shown in Fig. \ref{fig6} and nonlinear fitting $log(\sigma_M^2(s))=a+b(ds/s)+c(ds/s)^2$ . The standard errors of regression (\ref{E2}) against $x^2$ are denoted by an error bar. (B) $\sigma_M^2(s)=exp(a)$ (see the text). Standard errors of nonlinear fittings are denoted by an error bar. }
\label{fig8}
\end{center}
\end{figure}

\vspace{1cm}
\paragraph{$b_A$ and $\sigma_M$}
Similarly, we estimate parameters $b_A $ and $\sigma_M$ by taking the limit $\lambda_1-\lambda_2\rightarrow 0$ of the second moment $E((x_1-x_2)^2|x_2=x)/(\lambda_1-\lambda_2)$$(\ref{K-M})$.
The second moment $E((x_1-x_2)^2|x_2=x)/(\lambda_1-\lambda_2)$ is fitted by a quadratic function (a regression against $x^2$) as
\begin{equation}
\frac{E((x_1-x_2)^2|x_2=x)}{d\lambda}=b_A(\lambda_2,d\lambda)+\sigma_M(\lambda_2,d\lambda)x^2.
\label{E2}
\end{equation}
As shown in Fig. \ref{fig6}(a), the second moment is well fitted by a quadratic function. Fitting of this kind is applied to various $\lambda_1$ and $\lambda_2$ combinations (Fig. \ref{fig6}(B)). Taking the limit $d\lambda\rightarrow 0$, one obtains $b_A^2(\lambda_2)=\lim_{d\lambda\rightarrow  0}b_A^2(\lambda_2,d\lambda)$ and $\sigma_M^2=\lim_{d\lambda\rightarrow  0}\sigma_M^2(\lambda_2,d\lambda)$.
Fig. \ref{fig7}(A) presents examples of $b_A^2(s_2,ds/s)$ and nonlinear fitting by the function $log(b_A^2(s,ds/s))=a+b(ds/s)+c(ds/s)^2$. We estimate $b_A^2(s)$ for each line by $b_A^2(s)=\exp(a)$. The result is presented in Fig. \ref{fig7}(B). The solid line is the least-squares fit $b_A^2(s)$ to a power law function as
\begin{equation}
\log(b_A^2(s))=-1.67(0.56)+1.26(0.13)\log s,
\label{b_A}
\end{equation}
where the standard errors are in parentheses. The estimated exponent $1.26(0.13)$ is slightly higher than the constraint condition (\ref{assumption2})$(b_A^2(s)\sim s$). However, it is acceptable with accuracy. 
By a similar extrapolation $\log(\sigma_M^2(s,ds/s))=a+b(ds/s)+c(ds/s)^2$, we estimate $\sigma_M^2(s)=exp(a)$. Figure \ref{fig8}(A) presents an example of $\sigma_M^2(s_2,ds/s)$. and an estimate $\sigma_M^2(s)$ by $\sigma_M^2(s)=\exp(a)$ for each line. The result is shown in Fig. \ref{fig8}(B). We estimate parameter $\sigma_M^2$ by the average value weighted by the reciprocals of the standard errors.
\begin{equation}
\sigma_M^2=0.05(0.03)
\label{sig_M}
\end{equation}
Therein the standard error is in the parenthesis.

\begin{figure}[h!]
\begin{center}
\includegraphics[width=10cm]{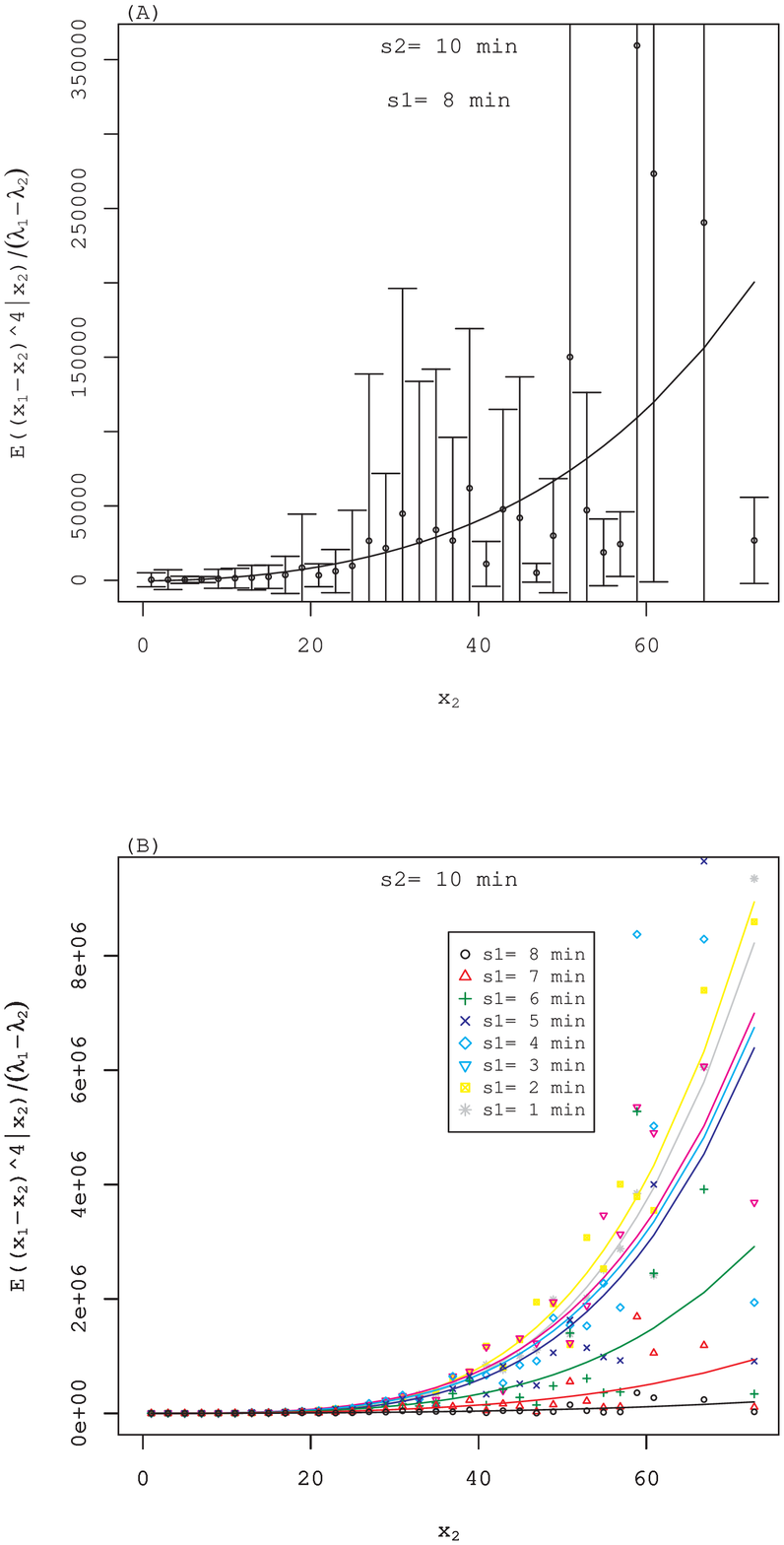}
\caption{Fitting of $E((x_1-x_2)^4|x_2=x)/(\lambda_1-\lambda_2)$ by a quartic function. (A) Standard errors are denoted by the error bar. (B) Fitting is applied to various $s_1$ and $s_2$ combinations.}
\label{fig9}
\end{center}
\end{figure}

\begin{figure}[h!]
\begin{center}
\includegraphics[width=10cm]{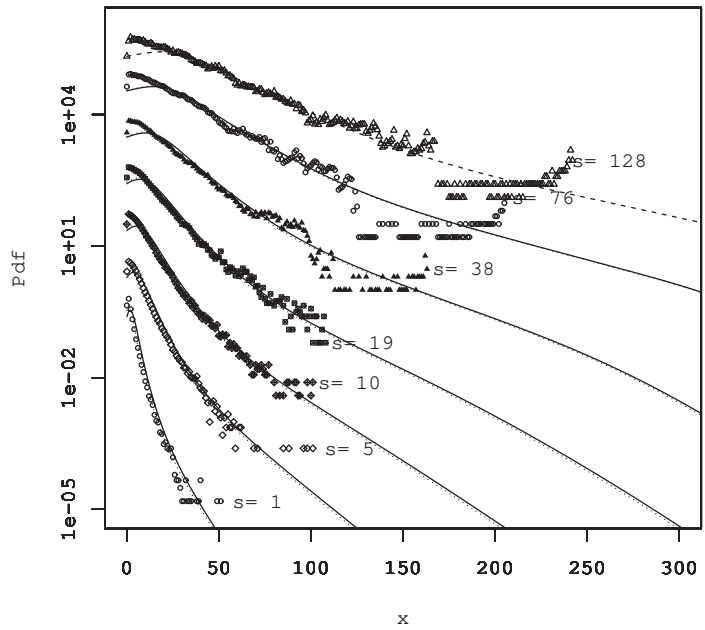}
\caption{The pdf of measured $x(\lambda)$ and numerical calculation of the Fokker--Planck equation. The result of numerical calculation is represented by the solid lines. Marks are measured values. The scale is attached to each line.}
\label{fig10}
\end{center}
\end{figure}

\vspace{1cm}
\paragraph{Higher moments}
Similarly, it is possible to show the k-th ($3 \le k$) moment $E((x_1-x_2)^k|x_2=x)/(\lambda_1-\lambda_2)$ of the transition probability density $p(x_1,\lambda_1|\lambda_2,x_2)$ vanishes in the limit $\lambda_1-\lambda_2\rightarrow 0$ .
As portrayed in Fig. \ref{fig9}(A), the fourth moment is well fitted by a quartic function. Applying the fitting to various $\lambda_1$ and $\lambda_2$ combinations (Fig. \ref{fig9}(B)). we have convinced that the fourth moment vanishes in the limit $d\lambda\rightarrow 0$. The Pawula theorem states that all higher Kramers--Moyal coefficients $D_k~(3 \le k)$ vanish if $D_4$ vanishes \cite{risken}. Therefore, we verified equation (\ref{KM}).

\vspace{1cm}
\paragraph{Numerical calculation of the Fokker--Planck equation}
We confirmed that estimation of the parameter $a_A(\lambda)$ and $b_A(\lambda)$ by the $E((x_1-x_2)^k|x_2=x)/(\lambda_1-\lambda_2~(k=1,2)$ is consistent with the constraint condition (\ref{assumption2}) with accuracy. If one imposes the other constraint (\ref{assumption3}), then the parameters have the following functional form of
\begin{align}
a_A(\lambda(s))&=\epsilon  s^{0.5}
\nonumber\\
b_A^2(\lambda(s))&=2.27\epsilon  s,
\label{abvalue}
\end{align}
where $\epsilon$ is a small parameter. The consistent range of $\epsilon$ found by estimation (\ref{a_A}) and (\ref{b_A}) is $0.15\le \epsilon \le 0.34$.
To fix parameter $\gamma_M$ and $\sigma_M$, we use the empirical value of the scaling exponent $\tau(q)$, which is fitted by the quadratic function $\tau(q)=-1+0.52q-0.013q^2$ (see Fig. \ref{fig2}(C)). One can derive $\tau(q)=-1+(\gamma_M+\frac 12\sigma_M^2)  q - \frac12 \sigma_M^2 q^2$ for the basic model (\ref{langevin1}) without additional stochastic processes. Again using the assumption of slight perturbation, then from the coefficients of the quadratic function, the parameters $\gamma_M$ and $\sigma_M$ are expected to exist respectively in the neighborhood of $\gamma_M=0.51$ and $\sigma_M=0.026$. 
Next we try the value of the parameters $\gamma_M=0.51$, $\sigma_M=0.026$ and $\epsilon=0.16$ for numerical calculation of the Fokker--Planck equation. 
Results are presented in Fig. \ref{fig10}. The initial pdf of the numerical calculation represented by the dashed line was based on the measured pdf on scale $s = 128 (min) $. In the initial values, the fine fluctuation was smoothed using a spline function with the rationale that small fluctuations in the measured pdf are attributable to the finiteness of the number of observations. The tails are extrapolated using a power function with index $-4.9 $ which is obtained empirically. For time evolution, the fourth-order explicit Runge--Kutta method was used. The solid line is the calculation result obtained using the estimated value of the parameters $\gamma_M=0.51$, $\sigma_M=0.026$ and $\epsilon=0.16$. The dotted line is the result obtained when $\epsilon=0$. The difference between the two was very small. The results closely matched the actual pdf.

\begin{figure}[h!]
\begin{center}
\includegraphics[width=10cm]{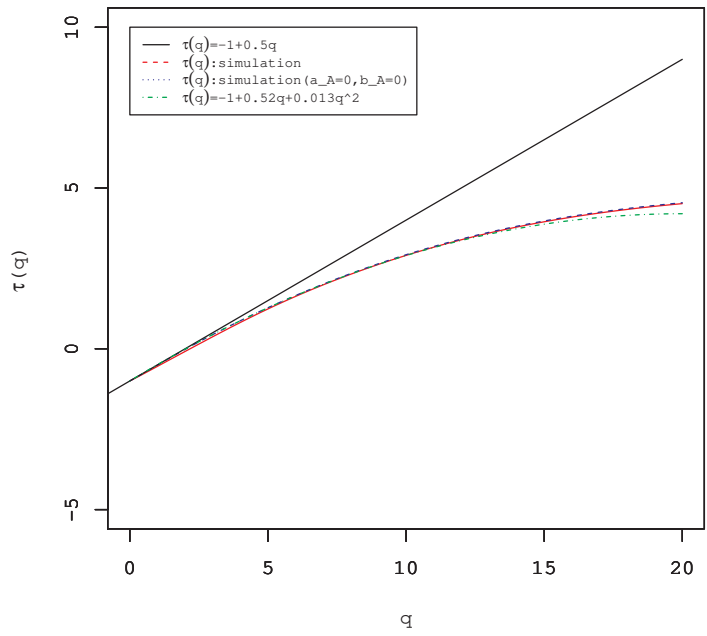}
\caption{Scaling exponent $\tau(q)$. A representation of each line is shown in the legend.}
\label{fig11}
\end{center}
\end{figure}

Using results of the numerical calculation of the pdf obtained at each scale, we calculate the scaling exponent $\tau(q)$ as shown below.\begin{equation}
E(|W_\psi Z[u,s]|^q)\sim s^{\tau(q)}~(0 \le q)
\label{ tauq2}
\end{equation}
The result is presented in Fig. \ref{fig11}. No difference exists between the two numerical calculation results. Both curves are convex upward, indicating multifractal properties. Comparison with measured values is also good. These results, when combined with consideration of the statistics of multipliers given in 2.1.2, underscore the effectiveness of the continuous cascade model (\ref{langevin2}) with additive stochastic processes proposed.

\vspace{1cm}
\section{Discussion}
The random cascade model has evolved as a model of developed turbulence. The original model, in which the stochastic process that connects each layer of the spatial scale is an independent random multiplication process, contradicts results obtained through empirical research. Therefore, an improved discrete random multiplicative cascade model with additional additive stochastic process was proposed along with a model formulated as a Fokker--Planck equation by considering cascade processes as a continuous Markov process. Moreover, those models have been applied to data analysis of the stock market and the foreign exchange market, where they have been successful. Herein, we propose a continuous cascade model formulated as a stochastic differential equation of volatility including two independent modes of Brownian motion: one has multiplicative coupling with volatility; the other has additive coupling, as in an improved discrete cascade model for the stock market, with effectiveness clarified by results of earlier research \cite{maskawa2018}. The model parameters were estimated by application to a stock price time series. The Fokker--Planck equation was derived from the stochastic differential equation as a master equation with the transition probability density function of volatility. Furthermore, the model parameters were estimated by its application to the average stock price time series made from FTSE100 constituents listed on the London Stock Exchange. At that time, as an alternative variable of volatility, the wavelet transform coefficient with the second derivative of the Gaussian function as an analyzing wavelet was used. Numerical calculation of the Fokker--Planck equation was conducted using the estimated parameter values. The results reported herein faithfully reproduce the results of an earlier empirical study. This model includes information about neither the time axis nor the sign of the price fluctuation, which are necessary for a model of price fluctuations. The actual stock market exhibits well known properties that break symmetry with respect to the time axis, such as the causal structure from long-term to short-term scale volatility described first in the Introduction and price--volatility correlation (Leverage effect) \cite{cont2001,bouchaud_potter}. Therefore the extension of the random cascade model to encompass these phenomena remains as a subject for future work.
 
\section*{Acknowledgments}
This research was partially supported by a Grant-in-Aid for Scientific Research (C) No. 16K01259.
One author, JM, expresses special appreciation for support by a Seijo University Special Research Grant.


\bibliographystyle{unsrt} 
\bibliography{arxiv_maskawa}


\section*{Figure captions}


\end{document}